\documentclass[aps,prb,twocolumn,superscriptaddress]{revtex4-2}

\usepackage{amsmath}
\usepackage{amssymb}
\usepackage{graphicx}
\usepackage{xcolor}
\usepackage{bm}
\usepackage{hyperref}

\begin{document}

\title{Finite-temperature spectral functions of lattice polarons with process tensors}

\date{\today}

\author{Qi Gao}
\affiliation{Institute of Physics, Chinese Academy of Sciences, Beijing 100190, China}

\author{Yuan Wan}
\email{yuan.wan@iphy.ac.cn}
\affiliation{Institute of Physics, Chinese Academy of Sciences, Beijing 100190, China}
\affiliation{Songshan Lake Materials Laboratory, Dongguan, Guangdong 523808, China}

\begin{abstract}
We develop a process tensor approach for calculating the finite-temperature spectral functions of lattice polarons. We show that the finite-temperature phonon influence functional can be expressed as the elementwise exponential of a low-rank matrix product state, enabling an efficient summation of electron worldlines. An embedding construction embeds worldlines of different propagation times into a single matrix product state. This construction yields the propagator over the full simulated time interval in a single calculation. We benchmark this approach against available finite-temperature density-matrix renormalization group results for the one-dimensional Holstein polaron. We further obtain numerically converged spectra for the two-dimensional Holstein polaron. The spectra reveal a progressive transfer of spectral weight from coherent spectral features into an incoherent background with increasing temperature.
\end{abstract}

\maketitle


\section{Introduction \label{sec:intro}}

The coupling between electrons and lattice vibrations shapes the electronic properties of solids~\cite{Ziman1960}. It influences charge transport, optical response, and the relaxation of electronic excitations in materials including transition-metal oxides~\cite{VanMechelen2008,Moser2013,Verdi2017}, organic semiconductors~\cite{Bredas1985,Hulea2006,Fetherolf2020}, and halide perovskites \cite{Wright2016,Herz2017,Miyata2017,Ghosh2020,Franchini2021}. A basic manifestation of this interaction is the polaron: a charge carrier accompanied by the lattice distortion that it induces~\cite{Landau1933,Frohlich1954,Holstein1959,Alexandrov2010}. The carrier and its surrounding phonon cloud move together as a dressed excitation, whose properties may differ substantially from those of the bare carrier. The single-particle spectral function of the polaron is directly accessible to spectroscopic probes, most notably angle-resolved photoemission \cite{Damascelli2003,Wang2016,Kang2018,Sobota2021}. By isolating the interaction of a single carrier with the lattice, the polaron problem provides a natural setting for understanding these effects and developing methods for electron-phonon systems~\cite{Jeckelmann1998,Prokofev1998}.

At finite temperature, the polaron moves through a lattice with thermally excited phonons~\cite{Holstein1959b}. Phonon absorption and emission modify the carrier's motion and redistribute its spectral weight. These modifications are manifest in the single-particle spectral function. The polaron lifetime is reduced, accompanied by a renormalization of its dispersion relation. Meanwhile, the absorption of thermally populated phonons by the charge carrier produces spectral weight below the zero-temperature ground-state energy \cite{Bonca2019}. At sufficiently high temperature, the polaron melts as its spectral function becomes incoherent \cite{Shannigrahi2026,Carbone2026}. Understanding how the polaron evolves with temperature, and under what conditions it loses coherence, remains a central question in polaron physics. 

The dynamical properties of a polaron at finite temperature have been explored using various numerical and analytical approaches~\cite{DeMello1997,Paganelli2006,Chen2017}. For the one-dimensional Holstein polaron model, Ref.~\onlinecite{Bonca2019} calculated the finite-temperature spectra using the finite-temperature Lanczos method~\cite{Jaklic1994} in a variationally constructed Hilbert space~\cite{Bonca1999}. Ref.~\onlinecite{Jansen2020} subsequently combined purification, time-dependent density-matrix renormalization group (DMRG), and local basis optimization~\cite{Zhang1998} to obtain spectra on substantially longer chains. The unfavorable scaling of the computational cost with dimensionality has restricted these methods largely to one-dimensional systems. These methods also require judiciously truncating the infinite-dimensional phonon Hilbert space without losing numerical accuracy, which may be difficult at high temperatures when the thermal population of phonons is large.  

Other numerical approaches make different trade-offs. Ref.~\onlinecite{Jankovic2022} applied hierarchical equations of motion to calculate the finite-temperature spectra on short one-dimensional chains. Free from phonon Hilbert space truncations, it converges rapidly at high temperature but incurs high computational cost as the temperature approaches zero. Ref.~\onlinecite{Mishchenko2015} obtained the finite-temperature polaron mobility using imaginary-time quantum Monte Carlo, which requires analytic continuation to recover real-frequency response. 

On the analytical front, various approximations have been used to tackle the lattice polaron problem at finite temperature. Dynamical mean-field theory approximates the electron-phonon interaction through a spatially local electron self-energy \cite{Ciuchi1997,Mitric2022}. More recent developments include the cumulant expansion~\cite{Robinson2022,Robinson2022b,Mitric2023}, the semiclassical method~\cite{Nguyen2025}, the finite-temperature generalization of the momentum-average approximation \cite{Berciu2006,Berciu2007,Shannigrahi2026}, the cluster expansion \cite{Carbone2021,Carbone2026}, and the variational coupled-cluster method \cite{Wu2026}.

Obtaining numerically exact finite-temperature polaron spectra beyond one dimension, therefore, remains an outstanding challenge. In this work, we establish a process tensor method for calculating the spectral function of lattice polarons at finite temperature. This method is numerically exact, free of Hilbert space truncation and analytic continuation, and applicable to higher dimensions. It generalizes our method \cite{Gao2025} for the zero-temperature lattice polaron problem to finite temperature.

In our previous work, we calculated the lattice polaron propagator through its real-time Feynman path integral representation~\cite{Feynman1955}. The phonons enter through an influence functional, which records how they respond to the electron's motion and act back on it. We show that the phonon influence functional can be cast exactly as an elementwise exponential of a low-rank matrix product state (MPS). Each MPS site corresponds to a time slice of the electron worldline. Utilizing this structure, we derive an exact flow equation for the polaron worldline amplitude that connects the noninteracting limit, corresponding to a free electron, to finite electron-phonon coupling. Solving the flow equation variationally produces an MPS approximation to the worldline amplitude. Contracting this MPS yields the numerically exact polaron propagator at zero temperature. Fourier transforming it yields the spectral function. This influence functional construction can be viewed as building a process tensor that encodes the influence of the phonon environment on the mobile impurity \cite{Keeling2026}.

\begin{figure*}[t]
\includegraphics[width = 0.8\textwidth]{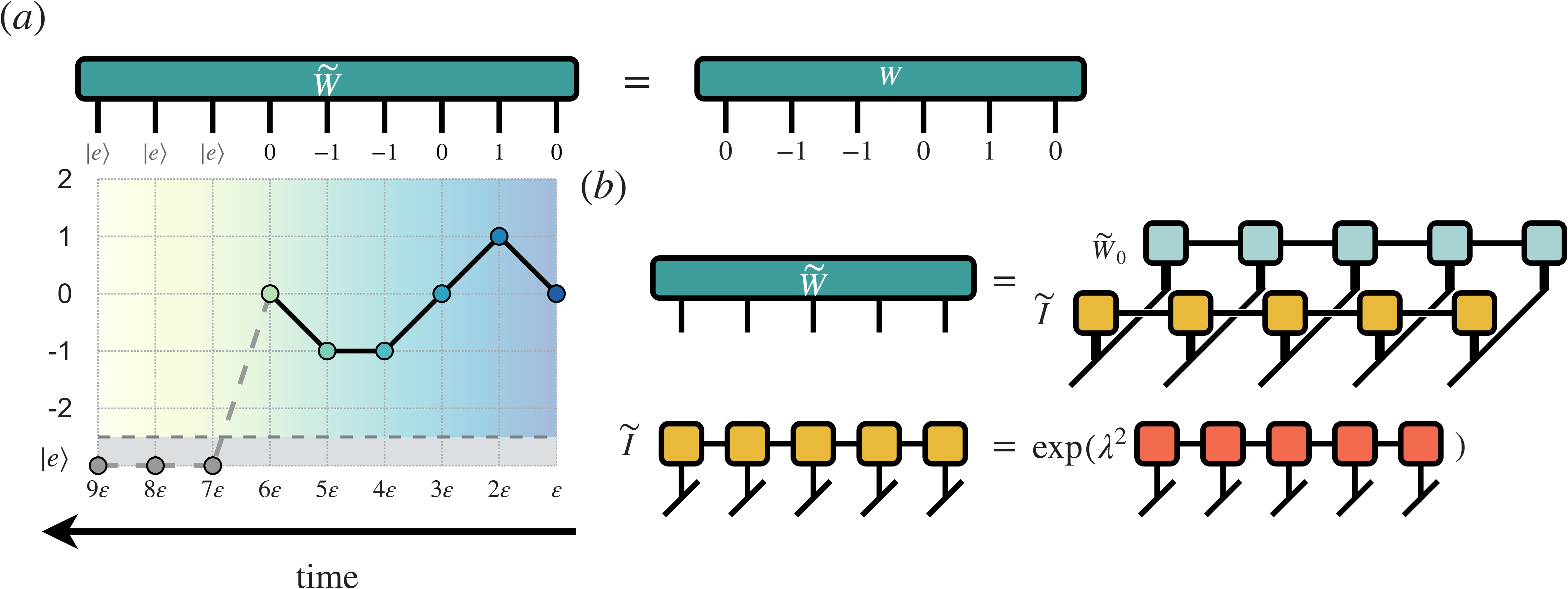}
\caption{Overview of the finite-temperature flow-equation approach. (a) The finite-temperature propagator is a sum over electron worldlines, parametrized by the displacements between successive Trotter slices. The worldline amplitude defines a high-dimensional tensor $\widetilde W$, which we approximate by a matrix product state. Slices in the extended ``empty'' state $|e\rangle$ (gray) represent late-time slices beyond the duration of the physical worldline. Thus, a single $\widetilde W$ contains the worldline amplitudes $W$ of all propagation times up to its length. Removing the empty slices reduces $\widetilde W$ to the amplitude $W$ of the physical worldline. (b) The amplitude factorizes into the free-electron amplitude $\widetilde W_0$ dressed by the phonon influence functional $\tilde I$, which is itself the exponential of a matrix product operator (MPO). $\widetilde W$ satisfies a flow equation in the coupling $\lambda^2$, which we integrate from $\lambda^2 = 0$ using the time-dependent variational principle. Throughout, tildes mark the embedded quantities of Sec.~\ref{sec:embedding}.}
\label{fig:overview}
\end{figure*}

The present work extends the process tensor construction for mobile impurities~\cite{Gao2025,Keeling2026} to finite temperature. The main difficulty lies in the thermal average over the phonons. We solve this problem by carrying out the average analytically in the phonon coherent state basis, which results in a Debye-Waller factor in the influence functional. This factor penalizes each worldline according to the magnitude of coherent phonon displacements. We show that the finite-temperature phonon influence functional can be exactly expressed as an elementwise exponential of a low-rank MPS similar to its zero-temperature counterpart. The temperature enters only as a parameter in this representation. We can therefore efficiently calculate the finite-temperature polaron propagator, and hence the spectral function, using essentially the same procedure as at zero temperature (Fig.~\ref{fig:overview}).

In the naive implementation of the process tensor approach, the length direction of the MPS corresponds to the physical time. Obtaining the complete time dependence of the polaron propagator requires multiple runs, each corresponding to one point on the time grid. We substantially reduce this computational cost by adapting an embedding construction~\cite{Link2024} that combines the separate calculations at different propagation times into a single MPS problem (Fig.~\ref{fig:embedding}). In other words, a single run produces the complete time dependence of the polaron propagator. 

We apply our method to the Holstein polaron in one and two dimensions. In one dimension, we obtain spectra in quantitative agreement with the available finite-temperature DMRG data~\cite{Jansen2020}, while extending the calculation to temperatures significantly higher than those reached by that method. In two dimensions, we demonstrate the convergence of our results with respect to system size and the MPS bond dimension, and benchmark the spectra against exact spectral sum rules. With increasing temperature, the spectra show a progressive suppression of coherent features and redistribution of their weight into an incoherent background. The quasiparticle peak can no longer be reliably separated from thermal features when the temperature reaches the phonon energy scale, indicating that the polaron has melted~\cite{Shannigrahi2026,Carbone2026}. 

Our work thus opens a numerically exact route to finite-temperature polaron dynamics in two dimensions. The rest of this paper is organized as follows. Sec.~\ref{sec:method} describes the process tensor approach to the polaron propagator at finite temperature and the embedding construction. Sec.~\ref{sec:results} benchmarks our method and presents the numerical results. Sec.~\ref{sec:discussion} discusses its relation to other process tensor methods and possible extensions.

\section{Method \label{sec:method}}

This section provides a detailed description of our method. We set up the problem in Sec.~\ref{sec:setup}. In Sec.~\ref{sec:flow_equation}, we recapitulate the zero-temperature flow-equation approach of Ref.~\onlinecite{Gao2025}. Sec.~\ref{sec:finite_T} contains the main analytical result of this work, namely the finite-temperature phonon influence functional. In Sec.~\ref{sec:carleman}, we show that the finite-temperature phonon influence functional can be represented as the elementwise exponential of a low-rank MPS using Carleman linearization. Finally, in Sec.~\ref{sec:embedding}, we present the embedding construction, which represents the propagator at all times up to a maximum time within a single MPS.

\subsection{The lattice polaron model \label{sec:setup}}

We consider a general lattice polaron Hamiltonian~\cite{Marchand2013},
\begin{subequations}
\label{eq:hamiltonian}
\begin{align}
H = K + V.
\end{align}
The first term,
\begin{align}
\label{eq:K}
K = -\sum_{\langle ij\rangle}\left(c^\dagger_i c^{\phantom\dagger}_j + \mathrm{h.c.}\right),
\end{align}
describes the nearest-neighbor hopping of a spinless electron on a periodic cluster of $N$ sites. $c_i$ ($c^\dagger_i$) annihilates (creates) an electron on lattice site $i$. The hopping amplitude is set to unity and serves as the unit of energy throughout this work. The second term,
\begin{align}
\label{eq:V}
V = \sum_{q}\left[\omega_q\, b^\dagger_q b^{\phantom\dagger}_q - \lambda f_q\, n_q \left(b^{\phantom\dagger}_q + b^\dagger_{-q}\right)\right],
\end{align}
describes the phonons and their coupling to the electron. $b_q$ ($b^\dagger_q$) annihilates (creates) a phonon with wavevector $q$. $\omega_q$ is the phonon dispersion. $\lambda>0$ is the coupling constant. $n_q = \sum_k c^\dagger_{k+q}c^{\phantom\dagger}_k/\sqrt{N}$ is the electron density in momentum space, where $N$ is the number of lattice sites. $f_q = f^\ast_{-q}$ is the form factor. The Holstein model corresponds to $\omega_q = \omega_0$ and $f_q = 1$.
\end{subequations}

We consider the retarded single-electron Green's function at temperature $T=1/\beta$,
\begin{equation}
\label{eq:propagator}
iG^R(k,t) =  \frac{1}{\mathcal{Z}}\mathrm{Tr}[e^{-\beta (H-\mu N_\mathrm{el})}\{c^{\phantom\dagger}_k(t),\, c^\dagger_k(0)\}].
\end{equation}
Throughout, we assume $t>0$. $\mu$ is the electron chemical potential. $N_\mathrm{el} = \sum_k c^\dagger_k c^{\phantom\dagger}_k$ is the total electron number. $\mathcal{Z} = \mathrm{Tr}(\exp(-\beta (H-\mu N_\mathrm{el})))$ is the grand canonical partition function.

We work in the limit $\mu\to-\infty$ relevant for an undoped system \cite{Bonca2019,Jansen2020}. In this limit, there is no electron to remove from the thermal ensemble, and $G^R$ reduces to the electron-addition Green's function. Moreover, the Boltzmann weight involves only the free-phonon part of the Hamiltonian, $H_0 = \sum_q \omega_q b^\dagger_q b^{\phantom\dagger}_q$, since the kinetic and electron-phonon coupling terms vanish in the absence of an electron. The propagator thus reduces to
\begin{align}
\label{eq:propagator_H0}
iG^R(k,t) = \frac{1}{\mathcal{Z}_0}\,\mathrm{Tr}\!\left[e^{-\beta H_0}\, e^{itH_0}\, c^{\phantom\dagger}_k\, e^{-itH}\, c^\dagger_k\right],
\end{align}
$\mathcal{Z}_0 = \mathrm{Tr}(e^{-\beta H_0})$ is the phonon partition function. In the limit $\beta\to\infty$, this expression reduces to the zero-temperature propagator of Ref.~\onlinecite{Gao2025}.

\subsection{The flow-equation approach \label{sec:flow_equation}}

Our method builds on the flow-equation approach developed for the zero-temperature problem in Ref.~\onlinecite{Gao2025}. We divide the propagation time $t$ into $M$ time slices of width $\varepsilon=t/M$. After applying a second-order Suzuki--Trotter decomposition to the incremental unitary evolution operator $\exp(-i\varepsilon H)$ and inserting electron position eigenstates at each time slice, the zero-temperature propagator becomes a sum over electron worldlines,
\begin{equation}
\label{eq:path_integral}
iG^R(k,t) = \sum_{\{x\}} e^{ikx_M}\, W_0(\{x\})\, I(\{x\}).
\end{equation}
$\{x\} = \{x_M,\cdots,x_1,x_0\}$ are the electron positions at successive time slices with the boundary condition $x_0 = 0$. The phase factor $\exp(ikx_M)$ comes from the Fourier transformation to momentum space. $W_0(\{x\}) = \prod_j \langle x_j|e^{-i\varepsilon K}|x_{j-1}\rangle$ is the worldline amplitude of a free electron. $I(\{x\})$ is the phonon influence functional,
\begin{equation}
\label{eq:I_x}
I(\{x\}) = \langle\phi|\, e^{-i\frac{\varepsilon}{2} V(x_M)}\cdots e^{-i\varepsilon V(x_1)}\, e^{-i \frac{\varepsilon}{2} V(x_0)}\,|\phi\rangle.
\end{equation}
$|\phi\rangle$ is the phonon vacuum. $V(x)$ is the phonon Hamiltonian with the electron pinned at site $x$:
\begin{align}
V(x) = \sum_q \left( \omega_q\, b^\dagger_q b^{\phantom\dagger}_q - \frac{\lambda f_q e^{iqx}}{\sqrt{N}}\, b^{\phantom\dagger}_q - \frac{\lambda f^\ast_q e^{-iqx}}{\sqrt{N}}\, b^\dagger_q \right).
\end{align}

It is convenient to parametrize the worldline by the displacements $s_j = x_j - x_{j-1}$ between successive time slices rather than the positions:
\begin{equation}
\label{eq:path_integral_s}
iG^R(k,t) = \sum_{\{s\}} \Big(\prod_{j=1}^{M} e^{iks_j}\Big)\, W_0(\{s\})\, I(\{s\}).
\end{equation}
The free amplitude then factorizes as $W_0(\{s\}) = \prod_j f(s_j)$, with $f(s) = \langle x+s|e^{-i\varepsilon K}|x\rangle$. We truncate the range of the displacement to $|s|\le 2$. The resulting error is of the same order as the Trotter error. On higher-dimensional lattices, the displacement simply becomes a lattice vector. The influence functional depends on the worldline through the displacement-controlled evolution operators
\begin{equation}
\label{eq:U_def}
U(s) = e^{-i\varepsilon V(0)/2} T(-s) e^{-i\varepsilon V(0)/2},
\end{equation}
where $T(a)$ is the phonon translation operator, translating the phonon degrees of freedom by $a$ lattice sites:
\begin{align}
\label{eq:T_action}
T(a)^{-1} b_q^{\phantom\dagger} T(a) = e^{-iqa} b_q^{\phantom\dagger}.
\end{align}
Using these, the phonon influence functional is cast in a compact form:
\begin{equation}
\label{eq:I_def}
I(\{s\}) = \langle\phi|\, T(r)\, U(s_M)\cdots U(s_1)\,|\phi\rangle .
\end{equation}
$r = \sum_j s_j$ is the total displacement of the worldline. At zero temperature, $T(r)$ can be omitted because it acts trivially on the translationally invariant vacuum. Here, we keep it explicit for the finite-temperature generalization in Sec.~\ref{sec:finite_T}.

A key observation of Ref.~\onlinecite{Gao2025} is that $I$ is the elementwise exponential of a low-rank MPS. To see this, we note that $U(s)$ maps the phonon coherent-state manifold onto itself. We parametrize a phonon coherent state as $\exp(\lambda^2 c)|\lambda \bm{z}\rangle$. $|\lambda \bm{z}\rangle = \exp(\sum_q \lambda z_q b^\dagger_q)|\phi\rangle$ is the canonical coherent state specified by the $N$ complex numbers $\bm{z} = (z_q)$. The complex parameter $c$ encodes the scalar prefactor of the phonon coherent state. For later convenience, they are respectively scaled by a factor of $\lambda$ and $\lambda^2$, where $\lambda$ is the electron-phonon coupling defined in Eq.~\eqref{eq:hamiltonian}.

As the vacuum $|\phi\rangle$ is a coherent state, the phonon state after $j$ time steps remains a coherent state. We parametrize this state as $ U(s_j)\cdots U(s_1)|\phi\rangle = \exp(\lambda^2 c_j)|\lambda \bm{z}_j\rangle$. The parameters obey a linear recursion,
\begin{equation}
\label{eq:coherent_recursion}
\begin{pmatrix} c_j \\ \bm{z}_j \\ 1 \end{pmatrix} = B_0(s_j) \begin{pmatrix} c_{j-1} \\ \bm{z}_{j-1} \\ 1 \end{pmatrix}.
\end{equation}
The initial condition is given by $c_0 = 0$, $\bm{z}_0 = \bm{0}$. $B_0(s)$ is an $(N+2)\times(N+2)$ transfer matrix that depends on the electron displacement $s$. It implements the action of $U(s)$ on the coherent-state parameters. The subscript $0$ emphasizes that this transfer matrix is for the zero-temperature problem. $B_0(s)$ is independent of the electron-phonon coupling $\lambda$ owing to the scaling of the coherent-state vector $\bm{z}$ and prefactor $c$. The explicit form of $B_0(s)$ is given in Appendix~\ref{app:transfer}.

Using the linear recursion relation Eq.~\eqref{eq:coherent_recursion}, we find the phonon influence functional takes the form,
\begin{align}
\label{eq:exponentiation}
I(\{s\}) = e^{\lambda^2 C(\{s\})}.
\end{align}
$C(\{s\})$ is the scalar coherent-state prefactor accumulated along the electron worldline,
\begin{align}
C(\{s\}) = v_{0,l}\, B_0(s_M)\cdots B_0(s_1)\, v_{0,r}.
\label{eq:C}
\end{align}
$C(\{s\})$ has an MPS representation with bond dimension $N+2$. The boundary vectors are $v_{0,r} = (0,\cdots,0,1)^{\mathsf T}$ and $v_{0,l} = (1,0,\cdots,0)$. Thus, Eq.~\eqref{eq:exponentiation} expresses the phonon influence functional as an elementwise exponential of a low-rank MPS.

The worldline amplitude $W(\{s\}) = W_0(\{s\}) I(\{s\})$ therefore obeys, configuration by configuration, the flow equation~\cite{Wegner1994,Gao2025},
\begin{equation}
\label{eq:flow_equation}
\frac{dW(\{s\})}{d\lambda^2} = C(\{s\})\, W(\{s\}),
\end{equation}
with the initial condition $W(\{s\})\big|_{\lambda^2=0} = W_0(\{s\})$. Although simple for each configuration, this differential equation is the key to our compression scheme. Formally, we view $W(\{s\})$ as a wave function $\langle s|W\rangle$ over worldline configurations, and $C(\{s\})$ as a diagonal operator $\hat C$, i.e. $\hat{C}|s\rangle = C(s)|s\rangle$. The flow equation then takes the form of an imaginary-time Schr\"odinger equation, $d|W\rangle/d\lambda^2 = \hat C|W\rangle$, with the coupling $\lambda^2$ playing the role of imaginary time and $\hat C$ playing that of a diagonal but non-Hermitian Hamiltonian (Fig.~\ref{fig:overview}b). This correspondence immediately suggests a compression scheme: the equation is solved variationally with MPS trial states of maximum bond dimension $\chi$ using the time-dependent variational principle (TDVP) \cite{Haegeman2011,Haegeman2016}. Increasing $\chi$ systematically improves the approximation.

Having obtained an MPS approximation to the worldline amplitude $W(\{s\}) = W_0(\{s\})I(\{s\})$ at a fixed bond dimension $\chi$, we calculate the electron propagator $G^R(k,t)$ by contracting the resulting MPS with the phase factors $\exp(iks_j)$ according to Eq.~\eqref{eq:path_integral_s}. A Fourier transform then yields the desired spectral function. Note that the momentum $k$ enters only at this final contraction. The momentum grid can be chosen much finer than the spacing $2\pi/N$ imposed by periodic boundary conditions, equivalent to imposing twisted boundary conditions. A finite system size $N$ primarily limits the spatial extent of the phonon cloud dressing the electron.

A polaron can be viewed as an impurity moving through a structured phonon environment, and our construction may accordingly be viewed as building the process tensor of this mobile impurity~\cite{Keeling2026}. The matrix-product influence functional Eq.~\ref{eq:exponentiation} encodes how the phonons, driven by the electron along its worldline, act back on it. The TDVP solution of the flow equation Eq.~\eqref{eq:flow_equation} yields  a low-bond-dimension MPS representation of the full worldline amplitude. The virtual space of this MPS is the dynamically active subspace of the full phonon Hilbert space.

The method summarized so far applies at zero temperature. Its finite-temperature extension, developed in Sec.~\ref{sec:finite_T}, retains the same framework. The path integral, the flow equation, and its TDVP solution all carry over verbatim. The thermal fluctuations only modify the form of the influence functional. At finite temperature, the phonon influence functional becomes $I_\beta = \exp(\lambda^2 C_\beta)$, where $C_\beta = C - \delta C_\beta$. $C$ denotes the zero-temperature scalar coherent-state prefactor Eq.~\eqref{eq:C}. $\delta C_\beta$ is the thermal correction derived in Sec.~\ref{sec:finite_T}.

\subsection{The finite-temperature influence functional \label{sec:finite_T}}

At finite temperature, the vacuum expectation value in $I(\{s\})$ is replaced by the thermal average over the free-phonon ensemble (Sec.~\ref{sec:setup}). The phonon influence functional is therefore given by
\begin{equation}
\label{eq:I_beta}
I_\beta(\{s\}) = \frac{1}{\mathcal{Z}_0}\,\mathrm{Tr}\!\left[ e^{(it-\beta)H_0}\, T(r)\, U(s_M)\cdots U(s_1) \right],
\end{equation}
where the trace is over the phonon Hilbert space. $H_0 = \sum_q \omega_q b^\dagger_q b^{\phantom\dagger}_q$ is the phonon Hamiltonian. $\mathcal{Z}_0 = \mathrm{Tr}\exp(-\beta H_0)$ is the partition function. 

Coherent states are the natural basis for evaluating this trace. To this end, we insert the resolution of identity $\bm{1} = \int_z \exp(-|\bm{z}|^2) |\bm{z}\rangle\langle\bm{z}|$. The integration measure is $\int \prod_q (d^2 z_q/\pi)$. The trace becomes an integral over the initial coherent-state amplitudes,
\begin{equation}
\label{eq:I_beta_integral}
I_\beta(\{s\}) = \frac{1}{\mathcal{Z}_0}\int_z e^{-|\bm{z}|^2}\,
\langle \bm{z}'|\, U(s_M)\cdots U(s_1)\,|\bm{z}\rangle .
\end{equation}
The bra $\langle\bm{z}'|$ is obtained by applying $\exp[(it-\beta)H_0]T(r)$ to $\langle \bm{z}|$,
\begin{equation}
\label{eq:z_prime}
z'_q = e^{-(\beta+it)\omega_q}\, e^{iqr}\, z_q.
\end{equation}

To evaluate the matrix element above, we consider the action of the unitary operator $U(s_M)\cdots U(s_1)$ acting on an arbitrary initial phonon coherent state $|\bm{z}\rangle$. The coherent-state recursion relation Eq.~\eqref{eq:coherent_recursion} gives
\begin{equation}
\label{eq:linear_action}
\begin{split}
U(s_M)\cdots U(s_1)\,|\bm{z}\rangle = {} & e^{\lambda^2 C(\{s\}) + X(\{s\})\,\bm{z}}\\
& \times |\lambda \bm{z}_M(\{s\}) + Y(\{s\})\,\bm{z}\rangle.
\end{split}
\end{equation}
The quantity $\exp(\lambda^2 C(\{s\}))|\lambda \bm{z}_M(\{s\})\rangle$ is the state obtained by evolving the phonon vacuum (Eq.~\eqref{eq:C}), which we have considered in Sec.~\ref{sec:flow_equation}. $\bm{z}_M(\{s\})$ is the final coherent-state vector, obtained by starting the recursion from the vacuum as discussed in Sec.~\ref{sec:flow_equation}. The row vector $X(\{s\})$ and the matrix $Y(\{s\})$ describe the corrections to the prefactor and the coherent-state vector, respectively, arising when the recursion starts from an arbitrary phonon coherent state $|\bm{z}\rangle$ rather than the phonon vacuum. The corrections are linear in the starting vector $\bm{z}$ because the recursion relation is linear.

The matrix $Y$ is diagonal,
\begin{align}
\label{eq:Y_mat}
(Y)_{qq'} = \delta_{qq'}\, e^{-i\omega_q t}\, e^{iqr}.
\end{align}
Its derivation is given in Appendix~\ref{app:Y}. $X(\{s\})$ and $Y(\{s\})$ are not independent:
\begin{equation}
\label{eq:X_identity}
X\,\bm{z} = -\lambda\, \bm{z}_M^\dagger\, Y\, \bm{z}.
\end{equation}
This relation follows directly from the unitarity of $U(s_M)\cdots U(s_1)$. Unitarity requires the norm of the evolved state to equal that of the initial state for every coherent state $|\bm{z}\rangle$, i.e. $2\,\mathrm{Re}\!\left(\lambda^2  C + X\bm{z}\right) + |\lambda \bm{z}_M + Y\bm{z}|^2 = |\bm{z}|^2$. Equating the terms linear in $\bm{z}$ yields the identity. 

We now calculate $I_\beta$. We substitute Eq.~\eqref{eq:linear_action} into the thermal average in Eq.~\eqref{eq:I_beta_integral}. Because the resulting integral is Gaussian, it can be evaluated analytically. Using Eqs.~\eqref{eq:z_prime}, \eqref{eq:Y_mat}, and \eqref{eq:X_identity}, we eliminate all other quantities in favor of $\bm{z}_M(\{s\})$: 
\begin{equation}
\label{eq:I_beta_result}
\begin{split}
I_\beta(\{s\}) &= e^{\lambda^2 C(\{s\})}\, \exp\!\left(-\lambda^2\sum_q \frac{|z_{Mq}(\{s\})|^2}{e^{\beta\omega_q}-1}\right)\\
&\equiv e^{\lambda^2\left[C(\{s\}) - \delta C_\beta(\{s\})\right]}.
\end{split}
\end{equation}
Temperature enters solely through a Debye-Waller factor. $\delta C_\beta$ encodes the correction due to phonon thermal fluctuations:
\begin{equation}
\label{eq:dC_beta}
\delta C_\beta(\{s\}) = \bm{z}_M^\dagger(\{s\})  \bm{D}\, \bm{z}_M(\{s\}).
\end{equation}
$(\bm{D})_{qq'} = \frac{\delta_{qq'}}{e^{\beta\omega_q}-1}$. $\delta C_\beta(\{s\})$ penalizes each worldline according to the magnitude of the displacement $|\bm{z}_M|$. As an immediate consequence, the zero-temperature construction is recovered in the limit $\beta\to\infty$.

\subsection{Matrix-product form through Carleman linearization \label{sec:carleman}}

In the previous section, we showed that the finite-temperature phonon influence functional can be expressed as $I_\beta = \exp(\lambda^2 C_\beta)$. In this section, we show that $C_\beta$ can be brought to a matrix-product form analogous to its zero-temperature counterpart. $C(\{s\})$ is already an MPS of bond dimension $N+2$ (Eq.~\eqref{eq:C}), and $\bm{z}_M(\{s\})$ is likewise an MPS of bond dimension $O(N)$. The thermal correction $\delta C_\beta$, however, is bilinear in $\bm{z}_M$. Thus, a direct substitution of the MPS form of $\bm{z}_M$ into $C_\beta$ would be computationally prohibitive.

We solve this problem using Carleman linearization \cite{Carleman1932}. In the zero-temperature case, we obtain the MPS representation of $C(\{s\})$ from the linear recursion for $\bm{z}_j$ and $c_j$ in Eq.~\eqref{eq:coherent_recursion}. The Carleman linearization promotes the bilinear quantity, $\bm{z}^\dagger_j \bm{D} \bm{z}_j$, to a component of an enlarged state vector. This enlarged state vector now admits a linear recursion relation,
\begin{align}
\begin{pmatrix} c_j \\ \bm{z}^\dagger_j \bm{D} \bm{z}_j \\ \bm{z}^\ast_j \\ \bm{z}_j \\ 1 \end{pmatrix}
=
B_\beta (s_j)\begin{pmatrix} c_{j-1} \\ \bm{z}^\dagger_{j-1} \bm{D} \bm{z}_{j-1} \\ \bm{z}^\ast_{j-1} \\ \bm{z}_{j-1} \\ 1 \end{pmatrix}.
\end{align}
The $(2N+3)\times (2N+3)$ transfer matrix $B_\beta(s)$ updates the coherent-state vector $\bm{z}$, its complex conjugate $\bm{z}^\ast$, and the bilinear form $\bm{z}^\dagger \bm{D} \bm{z}$. The subscript $\beta$ denotes the finite-temperature problem. We can read off the value of $\delta C_\beta = \bm{z}^\dagger_M \bm{D} \bm{z}_M$ from the second component of this enlarged state vector at the final time slice. We refer the reader to Appendix~\ref{app:transfer} for the explicit form of $B_\beta(s)$.

Using the above recursion relation, we find an MPS representation of $C_\beta$:
\begin{align}
\label{eq:C_beta}
C_\beta(\{s\}) & = C(\{s\}) - \delta C_\beta(\{s\})
\nonumber\\
& = v_{\beta,l}\, B_\beta (s_M)\cdots B_\beta (s_1)\, v_{\beta,r}.
\end{align}
The boundary vectors are $v_{\beta,r} = (0,0,0,0,1)^{\mathsf T}$ and $v_{\beta,l} = (1,-1,0,0,0)$. $C_\beta$ is an MPS of bond dimension $2N+3$, compared with the bond dimension $N+2$ of $C$. All other aspects are unchanged from the zero-temperature formulation. The flow equation Eq.~\eqref{eq:flow_equation}, therefore, applies as it stands, with $C_\beta$ in place of $C$. In particular, the inverse temperature enters only as a parameter of $C_\beta$.

Finally, we note that nothing in Secs.~\ref{sec:finite_T} and \ref{sec:carleman} is specific to one dimension. On a higher-dimensional lattice, the displacement variable becomes a lattice vector. The local physical dimension of the MPS changes accordingly \cite{Gao2025}. Neither the coherent-state recursion nor the Carleman linearization is affected. All two-dimensional results in Sec.~\ref{sec:results} are obtained with this straightforward generalization.

\subsection{The embedding construction \label{sec:embedding}}

\begin{figure*}[t]
\includegraphics[width = 0.75\textwidth]{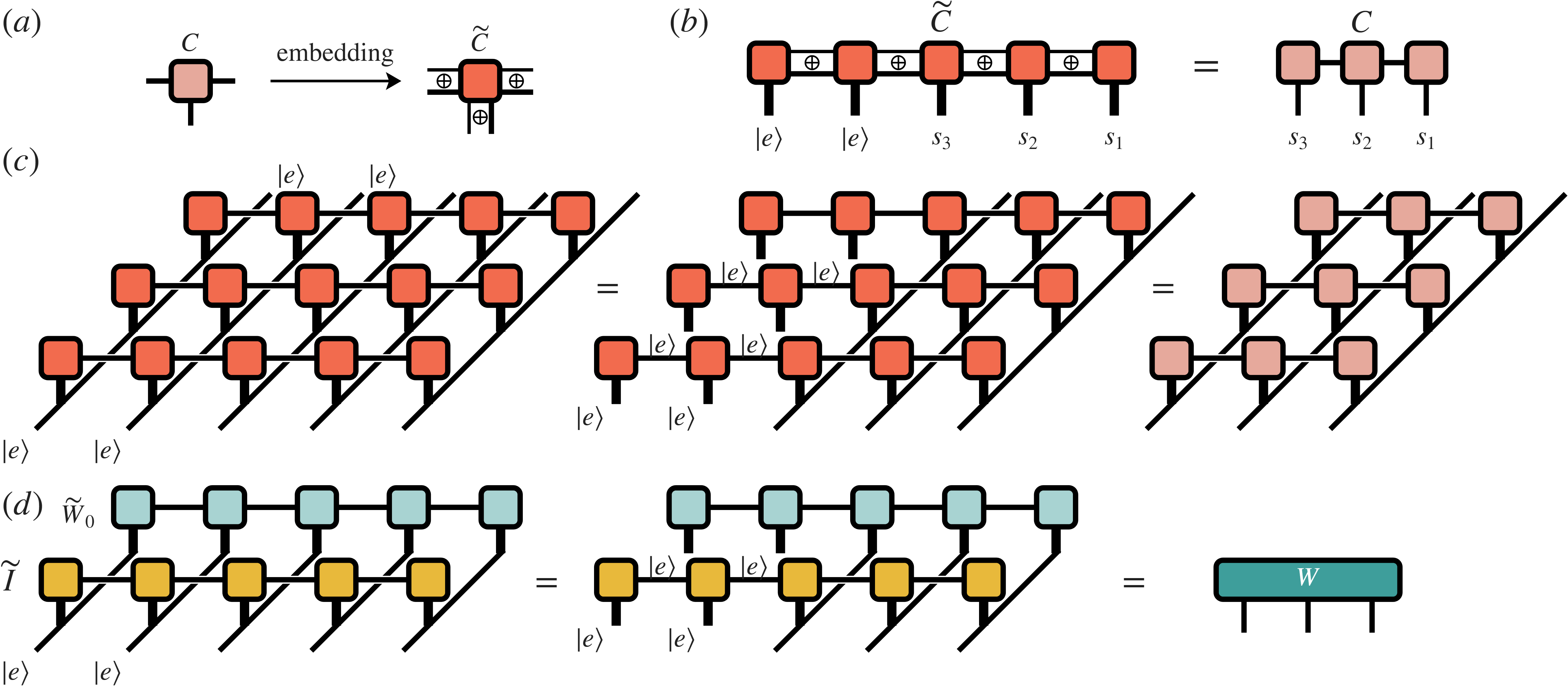}
\caption{Embedding construction. (a) At each site, the tensors of $C$ are promoted to those of $\widetilde C$ by a direct sum: both virtual legs are extended by one extra dimension, and the physical leg acquires the empty state $|e\rangle$. (b) Configurations of $\widetilde C$ whose late-time slices are set to $|e\rangle$ reduce to $C$ on the remaining slices. (c) This reduction is preserved under the matrix-product structure of the flow equation. The products of $\widetilde C$ reduce to the corresponding products of $C$. As a result, the influence functional obtained by exponentiation also reduces to that of the shorter worldline. (d) The reduction carries over to the amplitude. On an empty-terminated configuration, the elementwise product of $\tilde I$ and $\widetilde W_0$ yields the amplitude $W$ of the shorter worldline.}
\label{fig:embedding}
\end{figure*}

In the process tensor approach described above, physical time corresponds to the MPS length. To calculate the Fourier transform of the polaron propagator $G^R(k,t)$, we need its value on a time grid $t_j = j \varepsilon$, where $j = 0,1,2\cdots $ and $\varepsilon$ is the Trotter time step. As a result, for each point $t_j$ on the time grid, we need to run the flow equation compression over an MPS of length $L \propto j$. The maximal simulation time corresponds to a maximal MPS length $L_\mathrm{max}$. Obtaining $G^R(k,t)$ over this grid requires $O(L_\mathrm{max})$ TDVP runs.

We now embed the variable-length problem into a single fixed-length chain by adapting the extended tensor network construction of Ref.~\onlinecite{Link2024}. This construction enables calculation of $G^R(k,t)$ on the full time grid in a single TDVP run, thereby substantially reducing the computational cost.

The idea is to represent a worldline of length $L$ as a configuration of a longer chain of length $L_\mathrm{max}$, in which the first $L$ slices carry physical worldline displacement values, while the remaining $L_\mathrm{max}-L$ slices carry a new ``empty'' state $|e\rangle$. 

We first illustrate the embedding for $C(\{s\})$ (Eq.~\eqref{eq:C}). Generalization to $C_\beta$ is straightforward.  The physical index $\sigma$ is extended from the five allowed displacement values to six values, $\sigma \in \{e, \pm2, \pm1, 0\}$. The index $\sigma$ denotes the enlarged local state space. Meanwhile, we extend the virtual space by one dimension, labeled $0$ (Fig.~\ref{fig:embedding}a). The extended transfer matrices read
\begin{align}
\label{eq:tilde_B}
\tilde B_0(\sigma) = \left\{\begin{array}{ll}
\begin{pmatrix} 0 & \bm{0} \\ \bm{0} & B_0 (\sigma) \end{pmatrix} &  \sigma \in\{\pm 2,\pm 1,0\}
\\
\begin{pmatrix} 1 & v_{0,l} \\ \bm{0} & \bm{0} \end{pmatrix} & \sigma = e
\end{array}\right. .
\end{align}
We denote embedded quantities with a tilde. The boundary vectors become
\begin{equation}
\label{eq:tilde_boundaries}
\tilde v_{0,l} = \begin{pmatrix} 1 & v_{0,l} \end{pmatrix},
\qquad
\tilde v_{0,r} = \begin{pmatrix} 0 \\ v_{0,r} \end{pmatrix}.
\end{equation}
After embedding,
\begin{align}
\widetilde C(\{\sigma\}) = \tilde v_{0,l}\, \tilde B_0(\sigma_{L_\mathrm{max}})\cdots \tilde B_0(\sigma_1)\, \tilde v_{0,r}
\end{align} 
where time increases from right to left, as in Eq.~\eqref{eq:C}.

With these extended transfer matrices and boundary vectors, we find
\begin{align}
\label{eq:tilde_C}
\widetilde C(e,\cdots,e,\{s\}) = C(\{s\}).
\end{align}
The embedded MPS $\widetilde C$ of length $L_\mathrm{max}$ thus encodes worldlines of different lengths. Setting its final $L_\mathrm{max}-L$ time slices to $|e\rangle$ recovers $C$ for the physical worldline on the first $L$ slices (Fig.~\ref{fig:embedding}b). This reduction is preserved under elementwise powers, $\widetilde C^n\to C^n$ (Fig.~\ref{fig:embedding}c), and therefore holds term by term in the exponential expansion. The extended influence functional $\tilde I=\exp(\lambda^2\widetilde C)$ thus satisfies $\tilde I(e,\ldots,e,\{s\})=I(\{s\})$. For configurations in which a displacement slice occurs after an empty slice in time, $\widetilde C$ vanishes. A direct verification of these properties is given in Appendix~\ref{app:embedding}.

The remaining task is to find the embedded form of the initial amplitude $W_0(\{s\})$. We use an MPS of bond dimension two,
\begin{align}
\label{eq:W0_embedded_full}
\widetilde W_0(\{\sigma\}) = (1,1)\,A(\sigma_{L_\mathrm{max}})\cdots A(\sigma_1)\binom{0}{1},
\end{align}
where
\begin{equation}
\label{eq:W0_embedded}
A(s) = \begin{pmatrix} 0 & 0 \\ 0 & f(s) \end{pmatrix},
\quad
A(e) = \begin{pmatrix} 1 & 1 \\ 0 & 0 \end{pmatrix}.
\end{equation}
This construction satisfies $\widetilde W_0(e,\ldots,e,\{s\})=W_0(\{s\})$. As for $\widetilde{C}$, if a displacement slice occurs after an empty slice in time, $\widetilde W_0$ also vanishes.

Because both $\tilde I$ and $\widetilde W_0$ reduce to $I$ and $W_0$, respectively, on shorter worldlines, so does the full amplitude $\widetilde W=\tilde I\,\widetilde W_0$ (Fig.~\ref{fig:embedding}d). The embedded counterpart of the flow equation Eq.~\eqref{eq:flow_equation} is
\begin{equation}
\label{eq:embedded_flow}
\frac{d\widetilde W(\{\sigma\})}{d\lambda^2} = \widetilde C(\{\sigma\})\,\widetilde W(\{\sigma\}),
\end{equation}
with the initial condition $\widetilde W(\{\sigma\})\big|_{\lambda^2=0}=\widetilde W_0(\{\sigma\})$. At finite temperature, $\widetilde C$ is replaced by $\widetilde C_\beta$, obtained by embedding $C_\beta$.

Solving the embedded flow equation in a single TDVP run, we obtain the embedded worldline amplitude $\widetilde{W}$ that encodes all propagation times $t_j=j\varepsilon$ up to the maximal time. To extract $G^R(k,L\varepsilon)$, we fix the final $L_\mathrm{max}-L$ slices to $|e\rangle$ and contract the remaining $L$ physical indices with the phase factors $e^{iks_j}$ (Eq.~\eqref{eq:path_integral_s}). 

\section{Results \label{sec:results}}

In Sec.~\ref{sec:convergence}, we benchmark the method against finite-temperature reference data for the one-dimensional model. In two dimensions, we assess its accuracy through convergence tests and exact spectral sum rules. We then examine the temperature dependence of the polaron spectral function in both dimensions in Sec.~\ref{sec:temperature_dependence}.

\subsection{Benchmark \label{sec:convergence}}

\begin{figure}
\includegraphics[width = 0.9\columnwidth]{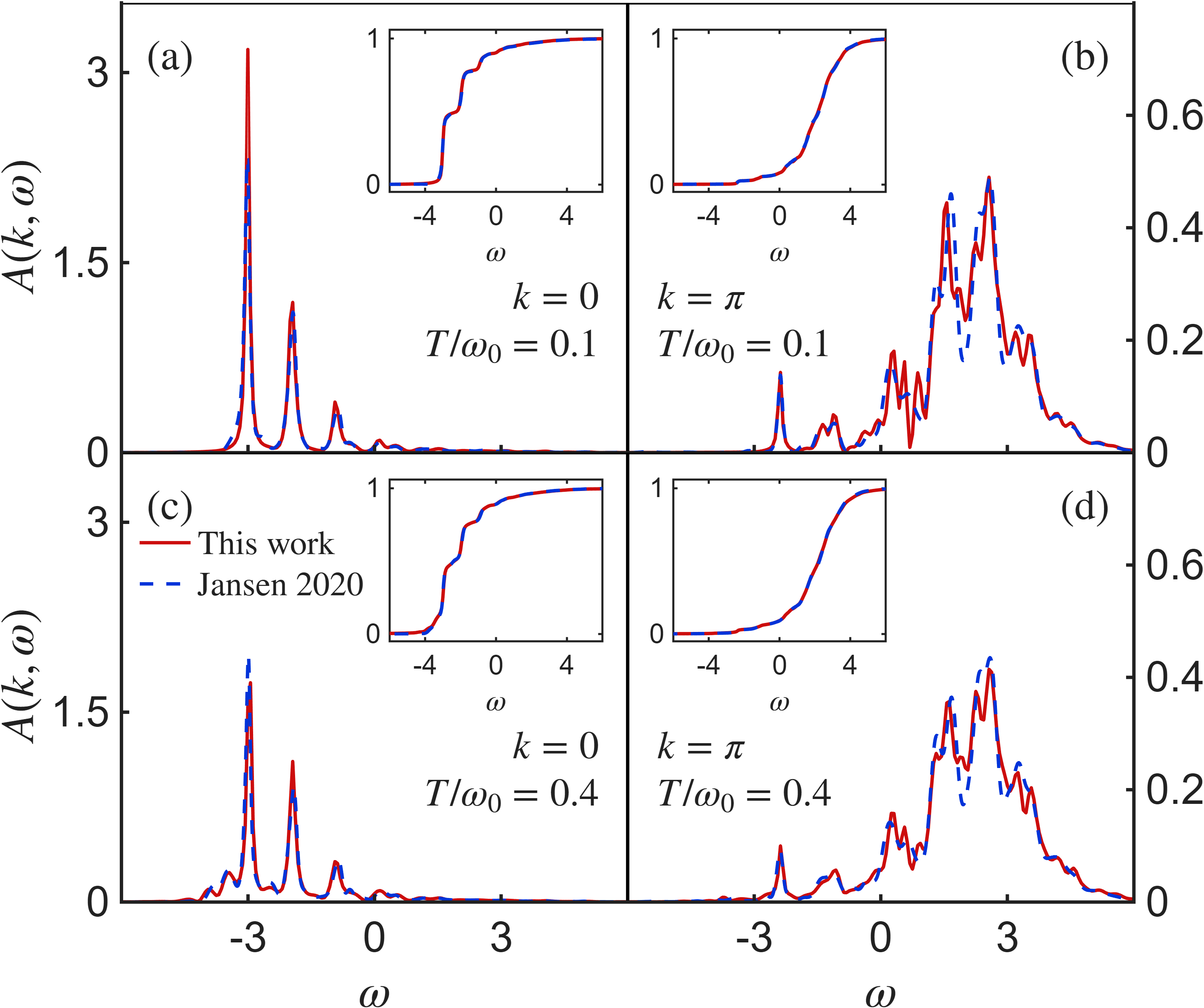}
\caption{Benchmark of the finite-temperature spectral functions of the one-dimensional Holstein polaron. Spectral functions $A(k,\omega)$ are shown for $\omega_q=\omega_0=1$, $f_q=1$, and $\lambda=\sqrt{2}$ at $k=0$ in panels (a) and (c) and at $k=\pi$ in panels (b) and (d). Panels (a) and (b) show $T/\omega_0=0.1$, whereas panels (c) and (d) show $T/\omega_0=0.4$. Red solid lines show results obtained on a chain of $N=24$ sites with maximum bond dimension $\chi=200$. Blue dashed lines show finite-temperature DMRG results of Ref.~\onlinecite{Jansen2020}. Insets show the integrated spectral function $\int_{-\infty}^{\omega}A(k,\omega')d\omega'$.}
\label{fig:benchmark_1d}
\end{figure}

A single TDVP evolution of the flow equation yields the retarded propagator $G^R(k,t)$ over the full simulated time interval $0\le t\le t_\mathrm{max}$ through the embedding construction described in Sec.~\ref{sec:embedding}. The momentum dependence is obtained from the final MPS contraction. We use a Trotter time step of $\varepsilon=0.1$ and evolve to $t_\mathrm{max}=20$ in one dimension; in two dimensions, we use $\varepsilon=0.05$ and $t_\mathrm{max}=10$. The spectral function is obtained from the Fourier transform of the retarded propagator:
\begin{align}
\label{eq:spectral_function}
A(k,\omega) &= -\frac{1}{\pi}\,\mathrm{Im}\,G^R(k,\omega),
\nonumber\\
G^R(k,\omega) &= \int_0^\infty G^R(k,t)\,e^{i\omega t}\,dt .
\end{align}
The Fourier transform requires the propagator for all $t\ge 0$. Following Refs.~\cite{White2008,Barthel2009}, we extend the computed time series using linear prediction before performing the transform. This improves the frequency resolution. We have checked that the peak positions and heights remain stable when the linear prediction parameters are varied. To suppress ringing artifacts and the associated negative spectral weight caused by the finite time window, we multiply the propagator by $e^{-\gamma t}$, with $\gamma=0.05$ in one dimension and $\gamma=0.4$ in two dimensions. This damping corresponds to adding a positive imaginary part to the frequency in the Lehmann representation and introduces a broadening set by $\gamma$.

Fig.~\ref{fig:benchmark_1d} compares our results for the one-dimensional Holstein polaron with the finite-temperature DMRG results of Ref.~\onlinecite{Jansen2020} at matched parameters and temperatures $T/\omega_0 = 0.1$ and $0.4$. We compare spectra at the Brillouin zone center ($k=0$) and boundary $(k=\pi)$. These momenta feature a prominent low-energy quasiparticle peak and a spectrum with strongly reduced quasiparticle weight, respectively. At $k=0$, both the main quasiparticle peak at $\omega/\omega_0 \approx -3.0$ and the phonon sidebands spaced by $\omega_0$ match quantitatively between the two methods. At $k=\pi$, the envelope and peak positions of the broad multiphonon continuum agree, while some fine structure of the incoherent background differs. These differences are much smaller in the integrated spectral function shown in the insets, which is less sensitive to the energy resolution and broadening procedure. This suggests that the differences mainly reflect the broadening and finite-time resolution used in the two calculations. Because the methods employ substantially different constructions, this agreement is reassuring.

\begin{figure}
\includegraphics[width = 0.9\columnwidth]{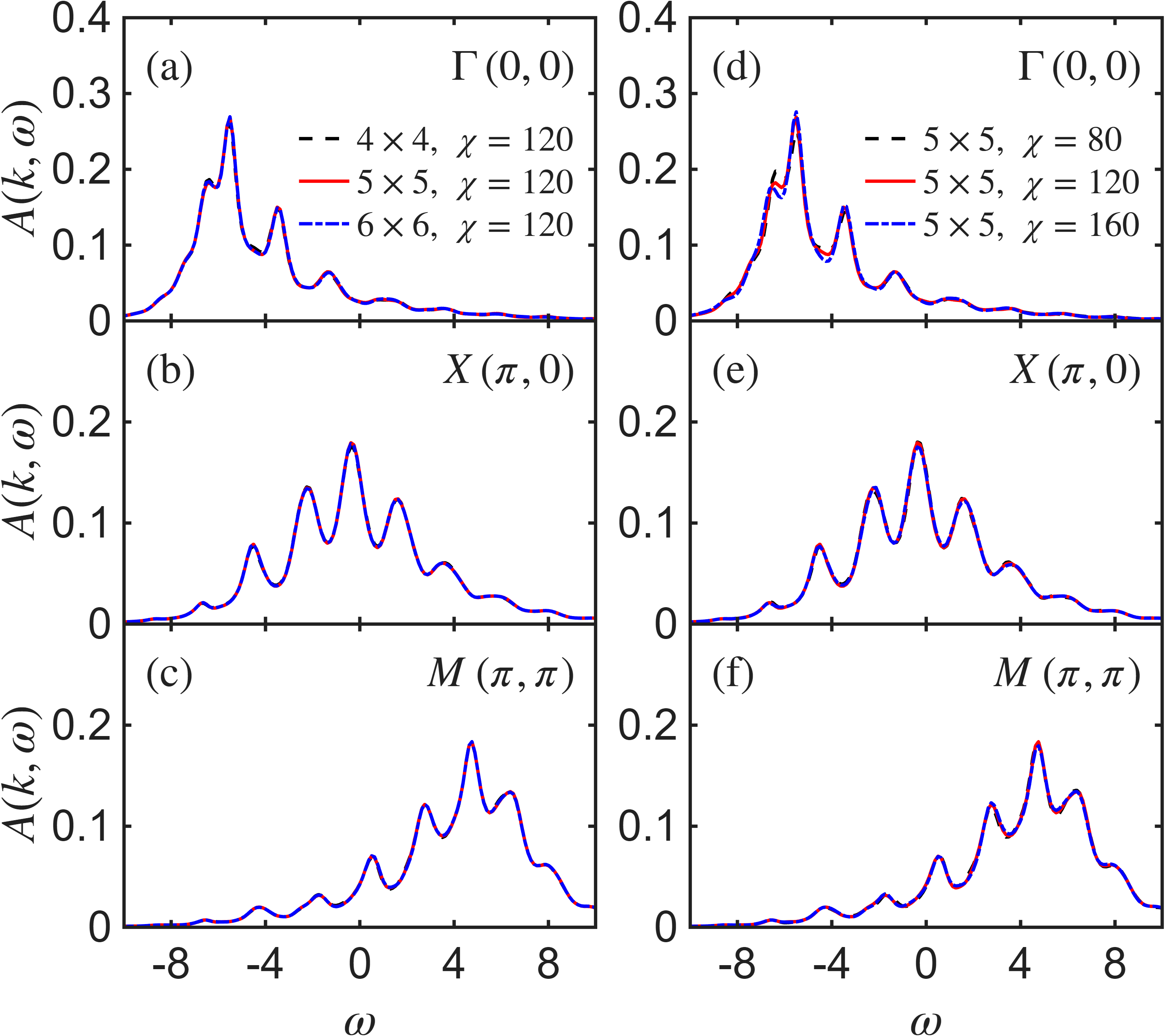}
\caption{Convergence of the spectral functions of the two-dimensional Holstein polaron. Spectral functions $A(k,\omega)$ are shown at the high-symmetry momenta $\Gamma=(0,0)$ in panels (a) and (d), $X=(\pi,0)$ in panels (b) and (e), and $M=(\pi,\pi)$ in panels (c) and (f), for $\omega_q=\omega_0=2$, $f_q=1$, $\lambda=2\sqrt{2}$, and $T=1.0$. The left column shows system sizes $4\times4$, $5\times5$, and $6\times6$ at $\chi=120$. The right column shows maximum bond dimensions $\chi=80$, $120$, and $160$ on the $5\times5$ cluster.}
\label{fig:convergence_2d}
\end{figure}

In two dimensions, to the best of our knowledge, no numerically exact finite-temperature spectral function data are available for the present model. We therefore validate our method internally. Fig.~\ref{fig:convergence_2d} shows $A(k,\omega)$ at the high-symmetry points $\Gamma$, $X$ and $M$ at the highest temperature considered, $T=1$, where the entanglement growth is more severe and convergence is more difficult to achieve compared with lower temperatures. The spectra are essentially unchanged when the system size increases from $4\times4$ to $6\times6$ at fixed $\chi=120$ (left column), as well as when the bond dimension increases from $\chi=80$ to $\chi=160$ on the $5\times5$ cluster (right column). Both the peak positions and the spectral weight distribution are robust. We conclude that a $6\times6$ cluster with $\chi=120$ provides converged results, and use it for the 2D data below.

\begin{figure}
\includegraphics[width = 0.9\columnwidth]{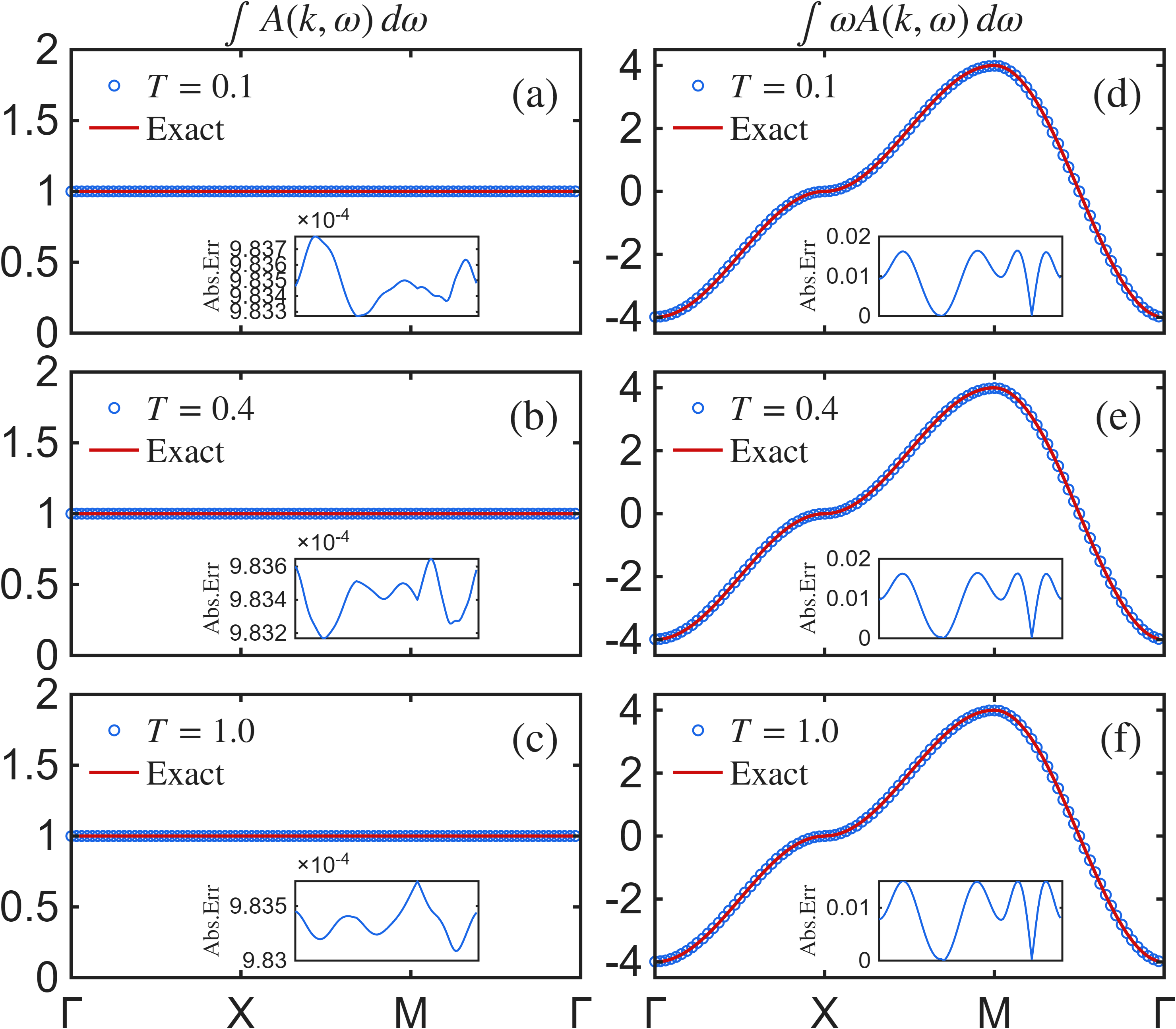}
\caption{Exact spectral sum rules for the spectral functions of the two-dimensional Holstein polaron. Zeroth moments are shown in panels (a)--(c), and first moments are shown in panels (d)--(f), along the $\Gamma$--$X$--$M$--$\Gamma$ path with 32 momentum points per segment. The rows correspond, from top to bottom, to $T=0.1$, $0.4$, and $1.0$, for a $6\times6$ cluster with $\chi=120$ and the model parameters of Fig.~\ref{fig:convergence_2d}. Blue open circles show numerical moments computed from the spectra, whereas red solid lines show the exact values. Insets show absolute errors along the momentum path.}
\label{fig:sum_rule}
\end{figure}

\begin{figure*}[t]
\includegraphics[width = 0.60\textwidth]{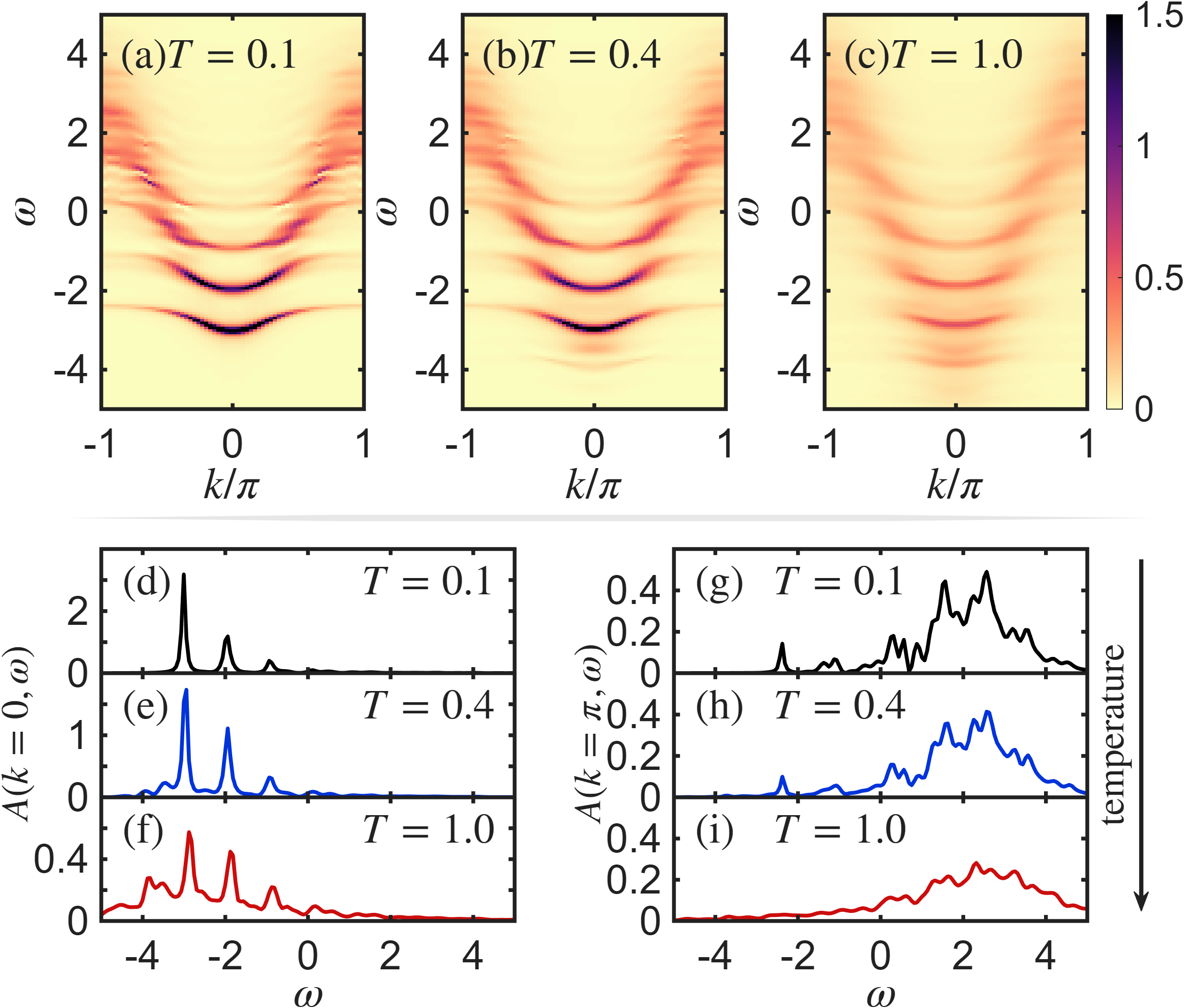}
\caption{Temperature dependence of the spectral functions of the one-dimensional Holstein polaron. Panels (a)--(c) show $A(k,\omega)$ for $\omega_0=1$, $\lambda=\sqrt{2}$, $N=24$, and $\chi=200$ at $T=0.1$, $0.4$, and $1.0$, respectively. The three heat maps share the same color scale. Panels (d)--(f) show constant-$k$ cuts at $k=0$, and panels (g)--(i) show cuts at $k=\pi$. Within both blocks, the temperature increases from top to bottom. The fixed-$k$ cuts share the same frequency axis. The $k=\pi$ panels share a common vertical scale, enabling direct comparison of the thermally suppressed quasiparticle peak.}
\label{fig:thermal_1d}
\end{figure*}

Beyond the internal convergence demonstrated above, we test the 2D spectra against exact sum rules \cite{Kornilovitch2002}
\begin{align}
\label{eq:sum_rules}
\int_{-\infty}^{\infty} A(k,\omega)\,d\omega &= 1,
\nonumber\\
\int_{-\infty}^{\infty} \omega\, A(k,\omega)\,d\omega &= E^{(0)}_k ,
\end{align}
where $E^{(0)}_k = -2(\cos k_x + \cos k_y)$ is the free-electron dispersion. Fig.~\ref{fig:sum_rule} shows the two moments evaluated from our spectra along the $\Gamma$--$X$--$M$--$\Gamma$ path. Both sum rules are satisfied at all temperatures considered, with absolute errors of order $10^{-3}$ for the zeroth moment and $10^{-2}$ for the first moment throughout the momentum path, as shown in the insets. The residual deviations most likely originate from Trotter discretization and the damping factor $\gamma$ used in the Fourier transform. These affect predominantly the high-energy part of the spectrum, to which the spectral moments are most sensitive.

\subsection{Temperature dependence of the spectral functions \label{sec:temperature_dependence}}

Fig.~\ref{fig:thermal_1d}(a)--(c) show the spectral functions for the Holstein polaron in one dimension at representative temperatures $T=0.1$, $0.4$ and $1$. We set the phonon frequency $\omega_0 = 1$ and the electron-phonon coupling $\lambda = \sqrt{2}$. We present our results for the system size $N = 24$ and the bond dimension $\chi = 200$, which are numerically converged, as demonstrated in Sec.~\ref{sec:convergence}.

At $T=0.1$ the spectrum closely resembles the zero-temperature one. A well-defined lowest polaron band with bandwidth strongly reduced compared to the free dispersion, and a comb of phonon sidebands separated by $\omega_0$ that bends upward away from $k=0$. Raising the temperature to $T=0.4$ broadens all features and transfers spectral weight from the coherent bands into the incoherent background. The sideband structure remains visible near $k=0$ but is largely washed out near the zone boundary. At $T=\omega_0 = 1$, the spectrum is dominated by a broad, weakly dispersive continuum. Only a remnant of the flattened polaron band near its bottom remains. Appendix~\ref{app:thermal_1d} presents spectral functions of the one-dimensional Holstein polaron from $T=0.01$ to $T=10$.

The constant-$k$ cuts provide a closer look at the temperature-dependent evolution of the spectral functions. At $k=0$ [Fig.~\ref{fig:thermal_1d}(d)--(f)], the quasiparticle peak broadens and acquires thermal satellite features on both low-energy and high-energy sides. At the zone boundary ($k=\pi$) [Fig.~\ref{fig:thermal_1d}(g)--(i)], the spectrum is already broad and multipeaked at $T=0.1$. With increasing temperature, the peaks flanking $\omega/\omega_0 \approx 2$ progressively lose spectral weight to the incoherent background, ultimately merging into a broad, nearly featureless continuum by $T=1$. Notably, the sharp peak near $\omega/\omega_0 \approx -2.5$, corresponding to the lowest polaron band near the zone boundary, is still visible at $T=0.4$ but disappears entirely by $T=1$, leaving no identifiable quasiparticle excitation at this momentum.

A distinctive feature of the finite-temperature spectra is the spectral weight that appears below the zero-temperature ground-state energy. It is already visible at $T=0.4$ and becomes increasingly pronounced at $T=1$ [Figs.~\ref{fig:thermal_1d}(e) and \ref{fig:thermal_1d}(f)]. This spectral weight originates from processes in which the electron absorbs a thermally excited phonon \cite{Bonca2019}. Such processes are forbidden at $T=0$.

\begin{figure*}[t]
\includegraphics[width = 0.9\textwidth]{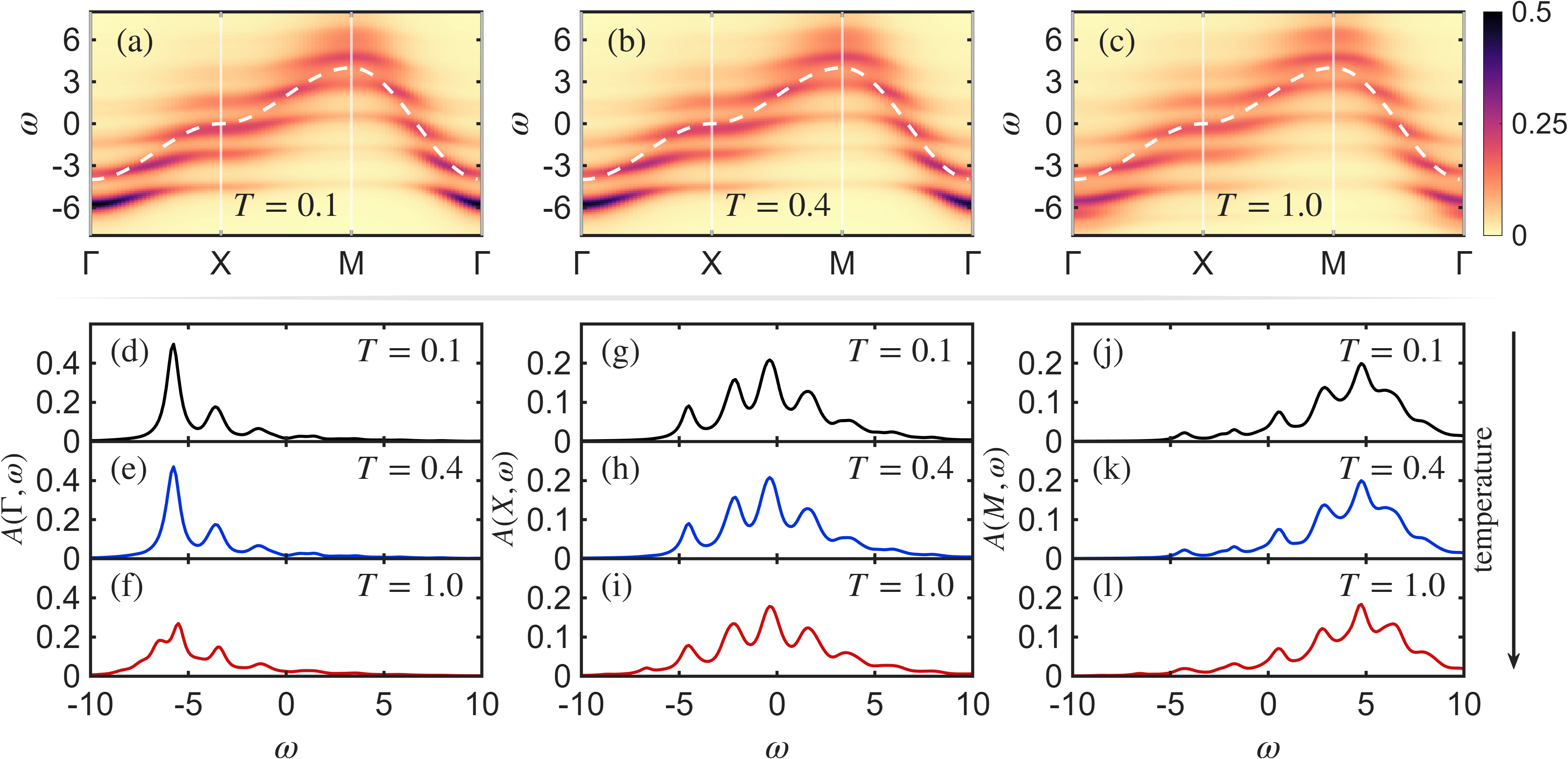}
\caption{Temperature dependence of the spectral functions of the two-dimensional Holstein polaron. Panels (a)--(c) show $A(k,\omega)$ along the $\Gamma$--$X$--$M$--$\Gamma$ path with 32 points per segment for $\omega_0=2$, $\lambda=2\sqrt{2}$, a $6\times6$ cluster, and $\chi=120$. The temperatures are $T=0.1$, $0.4$, and $1.0$ from left to right, corresponding to $T/\omega_0=0.05$, $0.2$, and $0.5$, respectively. White vertical lines mark the high-symmetry points, and the white dashed line shows the free-electron dispersion $E^{(0)}_k = -2(\cos k_x+\cos k_y)$. All panels share the same color scale. Panels (d)--(f), (g)--(i), and (j)--(l) show constant-$k$ cuts at $\Gamma$, $X$, and $M$, respectively. Within each block, the temperature increases from top to bottom, and the panels share the frequency axis and vertical scale.}
\label{fig:thermal_2d}
\end{figure*}

Overall, increasing the temperature progressively suppresses the coherent spectral features and transfers their weight to the incoherent background. This amounts to a thermal breakdown of the quasiparticle picture \cite{Shannigrahi2026,Carbone2026}. To quantify this process, we fit the $k=0$ spectra to multiple Lorentzian functions \cite{Shannigrahi2026}. We extract the polaron energy $E_P$, quasiparticle spectral weight $Z_T$, and thermal broadening $\eta^*(T) = \eta_T - \eta_0$, where $\eta_T$ denotes the linewidth at temperature $T$, and $\eta_0$ denotes the linewidth at $T=0.01$, used as the zero-temperature reference. We fit the quasiparticle peak at the lowest temperature with a single Lorentzian and the spectra at higher temperatures with a sum of four Lorentzians. We perform the fits using LMFIT \cite{lmfit}, and the results are summarized in Table~\ref{tab:qp_fit}.

\begin{table}[t]
\begin{ruledtabular}
\begin{tabular}{cccc}
$T$ & $E_P$ & $Z_T$ & $\eta^*$ \\
0.01 & $-3.00$ & $0.499$ & (baseline) \\
0.1 & $-3.00$ & $0.499$ & $+0.0002 \pm 0.0002$ \\
0.4 & $-2.96$ & $0.34$ & $+0.004 \pm 0.002$ \\
1.0 & --- & --- & --- \\
\end{tabular}
\end{ruledtabular}
\caption{Fitted quasiparticle parameters for the spectral functions of the one-dimensional Holstein polaron. The table lists the polaron energy $E_P$, quasiparticle spectral weight $Z_T$, and fitted excess linewidth $\eta^*$ at $k=0$ for the model parameters of Fig.~\ref{fig:thermal_1d}. These values are extracted from the multi-Lorentzian fits described in the text. The $T=0.01$ results provide the zero-temperature reference. At $T=1.0$, the quasiparticle peak cannot be reliably separated from thermal features, so no fitted parameters are reported.}
\label{tab:qp_fit}
\end{table}

The results at the two lowest temperatures are essentially indistinguishable. Between $T=0.01$ and $T=0.1$, the polaron energy and the quasiparticle weight are unchanged within numerical uncertainty, and the fitted thermal broadening $\eta^* = 0.0002 \pm 0.0002$ is statistically indistinguishable from zero. This behavior is expected since the thermal phonon populations at these temperatures are negligible.

Between $T=0.1$ and $T=0.4$, the polaron energy shifts upward by $\approx 0.04$, and the quasiparticle weight decreases from $Z_T \approx 0.50$ to $Z_T \approx 0.34$, which, using the relation $m^*/m_0 = 1/Z_T$, corresponds to an enhancement of the effective mass from $\approx 2.0$ to $\approx 2.9$. Meanwhile, a small but resolved thermal broadening $\eta^* = 0.004 \pm 0.002$ appears relative to the zero-temperature linewidth.

At $T=1$, however, multiple thermally induced features have comparable widths and spectral weights, preventing a reliable separation of the quasiparticle contribution from the incoherent background. We therefore report no fitted parameters at this temperature in Table~\ref{tab:qp_fit}. The quasiparticle picture is no longer well defined. Our results are consistent with those of Ref.~\onlinecite{Shannigrahi2026}.

Fig.~\ref{fig:thermal_2d} presents the polaron spectral function for the Holstein polaron on a two-dimensional square lattice at representative temperatures $T = 0.1$, $0.4$, and $1$. We set the phonon frequency $\omega_0 = 2$ and the electron-phonon coupling $\lambda = 2\sqrt{2}$. For this parameter set, the spectra are converged for a system of $6\times 6$ sites and the bond dimension $\chi = 120$.

The polaron spectral functions in two dimensions display the same overall temperature dependence as in one dimension. At $T=0.1$, we observe a well-defined polaron band whose bandwidth is significantly reduced relative to that of a free electron. The spectral weight of the higher bands shifts toward larger momenta, as in the zero-temperature case \cite{Gao2025}. For our present choice of phonon frequency $\omega_0=2$, the temperatures shown correspond to $T/\omega_0=0.05$, $0.2$, and $0.5$. Consequently, thermal smearing is milder than in the spectra shown in Fig.~\ref{fig:thermal_1d}.

At the $\Gamma$ point, the height of the quasiparticle peak changes only slightly as the temperature increases from $T=0.1$ to $0.4$, and decreases by approximately one-half at $T=1$. The spectra at $\Gamma$ exhibit a similar redistribution of spectral weight into a broad incoherent background as in the one-dimensional case [Fig.~\ref{fig:thermal_2d}(d)--(f)]. At $T=1$, the thermally activated weight below the zero-temperature ground-state energy discussed above is likewise visible at all three high-symmetry points [Figs.~\ref{fig:thermal_2d}(f), \ref{fig:thermal_2d}(i), and \ref{fig:thermal_2d}(l)]. Interestingly, at the X and M points, the redistribution of spectral weight is significantly less pronounced than at the $\Gamma$ point. The relative robustness of the spectral features at these momenta warrants further analytical investigation.

\section{Discussion \label{sec:discussion}}

To summarize, this work extends our previous zero-temperature process tensor approach to lattice polaron dynamics to finite temperature. We derive the finite-temperature phonon influence functional using the coherent-state representation and show, using Carleman linearization, that it is the elementwise exponential of a low-rank MPS. Meanwhile, our embedding construction makes it possible to map out the full time dependence of the polaron propagator in a single TDVP run. With these advances, we are able to obtain numerically converged finite-temperature spectra of Holstein polarons in one and two dimensions, supported by the benchmarks, convergence tests, and sum rule checks in Sec.~\ref{sec:results}.

Our method provides a starting point for polaron dynamics beyond the spectral function. A natural example is the spreading of an electron initially localized on one site characterized by its time-dependent density profile and mean-square displacement. Our earlier work \cite{Gao2025} formulated this problem as a path integral for the electron density and studied its zero-temperature evolution. Extending this calculation to finite temperature is now straightforward using the techniques developed here. Furthermore, it will be interesting to revisit the polaron mobility problem at finite temperature using our method, and compare the results with those from the quantum Monte Carlo and analytic continuation. Finally, although we have presented our results for one- and two-dimensional Holstein polarons, applying our method to three-dimensional models requires no modification.

Our method uses an interaction-flow equation that evolves the noninteracting electron worldline amplitude to that of a polaron at finite electron-phonon coupling $\lambda$. It is possible to derive a \emph{temperature}-flow equation that evolves the polaron at zero temperature to finite temperature. To see this, we note the worldline amplitude is given by $W = \exp(\lambda^2 C_\beta)W_0$. Taking the derivative with respect to the parameter temperature $T$ yields a differential equation: $\partial W/\partial T = -\lambda^2\beta^2 \partial_\beta C_\beta W$, where the quantity $\lambda^2 \beta^2 \partial_\beta C_\beta$ is the generator for the temperature flow. Crucially, it can be cast as a low-rank MPS as $C_\beta$ itself. This observation suggests that, starting from the amplitudes obtained at one reference temperature, we may obtain the amplitude across a range of temperatures by solving the temperature flow equation. Consequently, the results in Sec.~\ref{sec:results}, obtained from separate interaction-flow calculations at each temperature, could now be obtained by a single temperature flow run. Implementing this continuation and assessing its numerical convergence are left for future work.

The embedding described in Sec.~\ref{sec:embedding} is a general tensor network construction with applications beyond our process tensor method. It adapts the construction of \cite{Link2024}, which exploits the two-time influence-factor structure underlying the time-evolving matrix product operator (TEMPO) method. Our construction acts directly on local MPS/MPO tensors and their boundary vectors, without requiring a decomposition into two-time influence factors. In particular, the chain-length index can encode spatial size as well as propagation time. For example, real-time TDVP could evolve spin chains of different lengths after a quench from a product state within a single MPS calculation. This would provide access to the size dependence of nonequilibrium dynamics and facilitate the analysis of finite-size effects.

The difference between our approach and the established open quantum system methods such as TEMPO \cite{Strathearn2018,Jorgensen2019} lies in how the matrix-product influence functional is constructed. TEMPO constructs it by contracting and compressing a network of analytically known two-time influence factors at fixed coupling strength. The bond dimension of this network scales with the Hilbert space dimension of the physical system. For a mobile impurity, however, this dimension is $N$ as the mobile impurity can occupy $N$ positions, making a direct application of TEMPO costly \cite{Keeling2026}. Our approach avoids this issue by parametrizing the impurity worldline by its successive displacements, and constructing the MPS representation of the worldline amplitude by evolving it from the noninteracting limit to the desired coupling strength.

The mobile impurity viewpoint also suggests extensions to Bose polarons \cite{Grusdt2024}, Fermi polarons \cite{Massignan2014}, and spin polarons \cite{Bohrdt2021}. For Bose polarons, an enlarged manifold including squeezed coherent states could capture additional correlations of the surrounding medium. Fermionic and spin environments require influence functionals suited to their statistics and interactions. Tensor cross interpolation \cite{NunezFernandez2025} offers a possible route to compact tensor representations when the coherent-state recursion is unavailable. Combining these representations with the embedding construction could broaden the scope of our method to a broad class of dynamical problems involving mobile impurities.

\begin{acknowledgments}
We thank Honghao Tu for helpful discussions on the embedding construction. This work is supported by the National Key R\&D Program of China (Grants No. 2024YFA1408700, 2022YFA1403800), the Natural Science Foundation of China (Grants No. 12547112, 12188101), the CAS Project for Young Scientists in Basic Research (Grant No. YSBR-059).
\end{acknowledgments}

\appendix

\begin{figure*}[t]
\includegraphics[width = 0.8\textwidth]{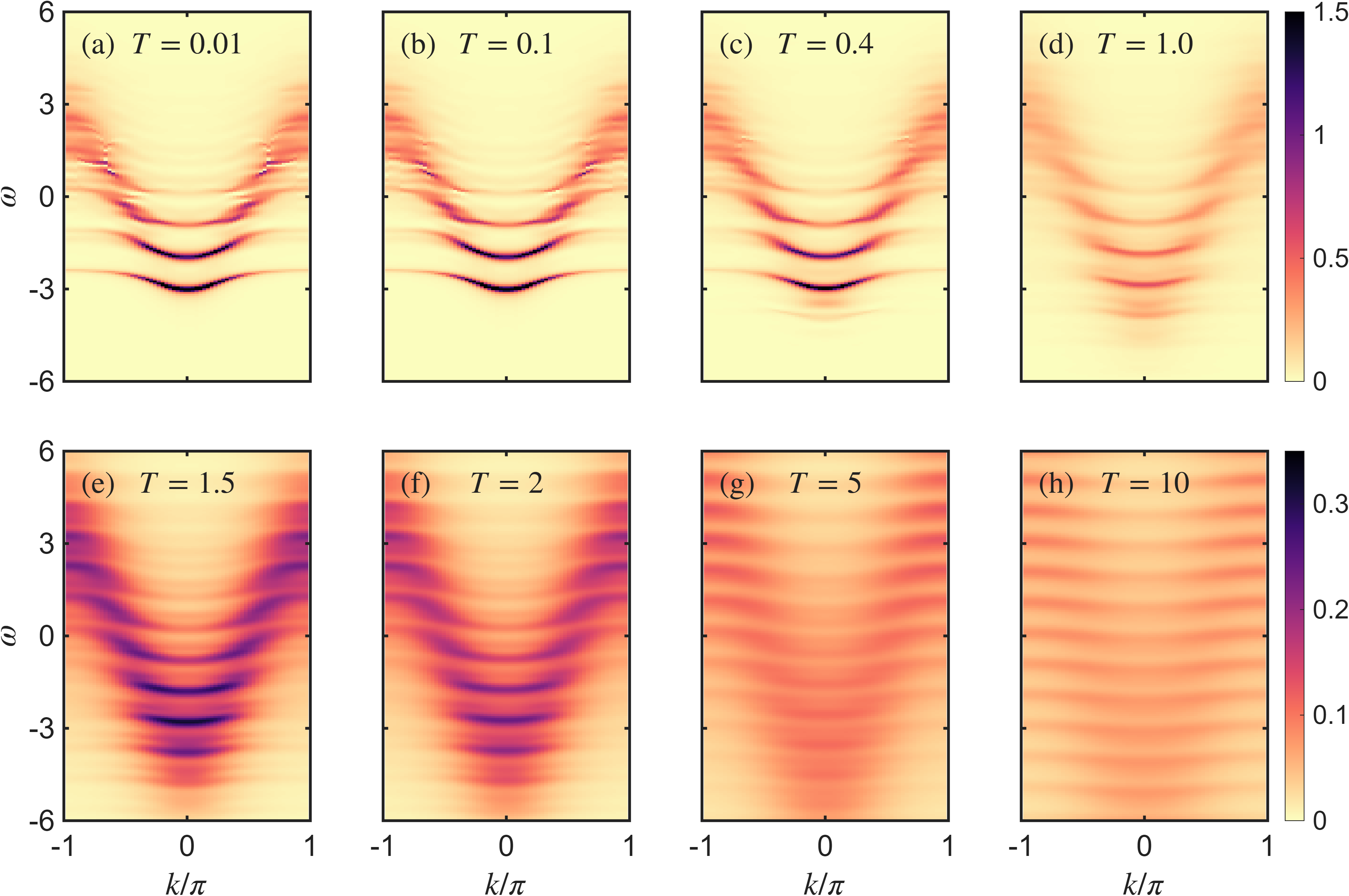}
\caption{Spectral functions of the one-dimensional Holstein polaron over an extended temperature range. The model parameters are $\omega_0=1$, $\lambda=\sqrt{2}$, $N=24$, and $\chi=200$. Panels (a)--(h) correspond, respectively, to $T=0.01$, $0.1$, $0.4$, $1.0$, $1.5$, $2$, $5$, and $10$. Each row shares a color scale, but the two rows use different color scales.}
\label{fig:thermal_1d_appendix}
\end{figure*}

\section{Explicit form of the transfer matrices \label{app:transfer}}

We give the explicit transfer matrices $B_0(s)$ and $B_\beta(s)$ used in Secs.~\ref{sec:flow_equation} and \ref{sec:carleman}. For the block ordering $(c,\bm{z},1)^{\mathsf T}$ in Eq.~\eqref{eq:coherent_recursion}, the zero-temperature transfer matrix is
\begin{equation}
\label{eq:transfer_matrices}
\begin{aligned}
B_0(s) &= B_{0,V} B_{0,T}(s) B_{0,V},\\
B_{0,V} &= \begin{pmatrix}
1 & -\bm{\alpha}^\dagger\bm{\Gamma} & \delta \\
0 & \bm{\Gamma} & \bm{\alpha} \\
0 & 0 & 1
\end{pmatrix},\\
B_{0,T}(s) &= \operatorname{diag}\!\left(1,\bm{\Delta}(s),1\right).
\end{aligned}
\end{equation}
The factors $B_{0,V}$ and $B_{0,T}(s)$ represent the half-step evolution $e^{-i\varepsilon V(0)/2}$ and the translation $T(-s)$, respectively, in the order of Eq.~\eqref{eq:U_def}. Here, the diagonal matrices $\bm{\Gamma}$ and $\bm{\Delta}(s)$, the vector $\bm{\alpha}$, and the scalar $\delta$ are defined by
\begin{equation}
\label{eq:app_transfer_entries}
\begin{aligned}
(\bm{\Gamma})_{qq'}
&=\delta_{qq'}e^{-i\omega_q\varepsilon/2},\\
(\bm{\Delta}(s))_{qq'}
&=\delta_{qq'}e^{iqs},\\
\alpha_q
&=\frac{f_q^\ast}{\sqrt{N}\omega_q}
\big(1-e^{-i\omega_q\varepsilon/2}\big),\\
\delta
&=\sum_q\frac{|f_q|^2}{N\omega_q}
\left[\frac{i\varepsilon}{2}
-\frac{1-e^{-i\omega_q\varepsilon/2}}{\omega_q}\right].
\end{aligned}
\end{equation}
The coherent-state derivation is given in \cite{Gao2025}.

At finite temperature, the block ordering is $(c,\bm{z}^\dagger\bm{D}\bm{z},\bm{z}^\ast,\bm{z},1)^{\mathsf T}$, with block sizes $1\oplus1\oplus N\oplus N\oplus1$, as in Sec.~\ref{sec:carleman}. The transfer matrix reads
\begin{equation}
\label{eq:app_B_beta}
B_\beta(s)=B_{\beta,V}B_{\beta,T}(s)B_{\beta,V},
\end{equation}
where
\begin{equation}
\label{eq:carleman_matrices}
B_{\beta,V}=\setlength{\arraycolsep}{3pt}\begin{pmatrix}
1 & 0 & 0 & -\bm{\alpha}^\dagger\bm{\Gamma} & \delta \\
0 & 1 & \bm{\alpha}^{\mathsf T}\bm{\Gamma}^\ast\bm{D} & \bm{\alpha}^\dagger\bm{\Gamma}\bm{D} & \bm{\alpha}^\dagger\bm{D}\bm{\alpha} \\
0 & 0 & \bm{\Gamma}^\ast & 0 & \bm{\alpha}^\ast \\
0 & 0 & 0 & \bm{\Gamma} & \bm{\alpha} \\
0 & 0 & 0 & 0 & 1
\end{pmatrix},
\end{equation}
\begin{equation}
\label{eq:app_B_beta_translation}
B_{\beta,T}(s)=\operatorname{diag}\!\left(1,1,\bm{\Delta}^\ast(s),\bm{\Delta}(s),1\right).
\end{equation}
Here, $\bm{D}$ is the diagonal matrix of Bose occupation numbers defined in Sec.~\ref{sec:finite_T}. All entries are independent of $\lambda$. Retaining the first, fourth, and fifth blocks recovers $B_0(s)$.

\section{Derivation of the matrix \texorpdfstring{$Y$}{Y} \label{app:Y}}

The transfer matrix $B_0(s_j)$ in Eq.~\eqref{eq:transfer_matrices} gives the coherent-state vector update
\begin{equation}
\label{eq:app_Y_recursion}
\bm{z}_j
=\bm{\Gamma}\bm{\Delta}(s_j)\bm{\Gamma}\bm{z}_{j-1}
+\bm{\Gamma}\bm{\Delta}(s_j)\bm{\alpha}+\bm{\alpha}.
\end{equation}
The last two terms are independent of the initial coherent-state vector. The contribution of the initial coherent-state vector to the final vector is therefore propagated by the product
\begin{equation}
\label{eq:app_Y_product}
Y(\{s\})=
\big[\bm{\Gamma}\bm{\Delta}(s_M)\bm{\Gamma}\big]\cdots
\big[\bm{\Gamma}\bm{\Delta}(s_1)\bm{\Gamma}\big].
\end{equation}

The diagonal entries of $\bm{\Gamma}$ and $\bm{\Delta}(s_j)$ in Eq.~\eqref{eq:app_transfer_entries} then give
\begin{equation}
\label{eq:app_Y_diagonal}
\begin{aligned}
(Y(\{s\}))_{qq'}
&=\delta_{qq'}\prod_{j=1}^{M}
\big(e^{-i\omega_q\varepsilon}e^{iqs_j}\big)\\
&=\delta_{qq'}e^{-i\omega_q t}e^{iqr},
\end{aligned}
\end{equation}
where $t=M\varepsilon$ and $r=\sum_{j=1}^{M}s_j$.

\section{Embedding properties \label{app:embedding}}

We demonstrate the embedding properties of the embedded quantities $\widetilde{C}$ and $\widetilde{W}_0$ defined in Sec.~\ref{sec:embedding}. These properties follow directly from the embedded tensors and boundary vectors.

We first demonstrate the embedding property for $\widetilde{C}$. Eqs.~\eqref{eq:tilde_B} and \eqref{eq:tilde_boundaries} give
\begin{equation}
\tilde v_{0,l}\,\tilde B_0(e)=\tilde v_{0,l},
\quad \tilde B_0(s)\,\tilde B_0(e)=0.
\end{equation}
The first identity implies that
\begin{align}
\tilde v_{0,l} \widetilde{B}_0(e)\cdots \widetilde{B}_0(e) \tilde B_0 (s_L)\cdots \tilde B_0 (s_1)\, \tilde v_{0,r}
\nonumber\\
 =v_{0,l}\,B_0(s_L)\cdots B_0(s_1)\,v_{0,r}=C(\{s\}),
\end{align}
which proves Eq.~\eqref{eq:tilde_C}.  Meanwhile, a configuration in which a nonempty slice occurs later in time than an empty slice contains the adjacent factor $\tilde B_0(s)\tilde B_0(e)$. The second identity therefore makes $\widetilde C$ vanish on such configurations. These identities also hold for $B_\beta(s)$ and the corresponding boundary vectors in Sec.~\ref{sec:carleman}.

The same reasoning applies to the extended free amplitude $\widetilde{W}_0$. We have:
\begin{equation}
(1,1)\,A(e)=(1,1),\quad A(s) A(e)=0.
\end{equation}
It follows that
\begin{align}
& (1,1)\, A(e)\cdots A(e)A(s_L)\cdots A(s_1)\binom{0}{1} 
\nonumber\\
& \quad = (1,1)\,A(s_L)\cdots A(s_1)\binom{0}{1} = W_0(\{s\}).
\end{align}
Any configuration in which an empty slice is followed in time by a nonempty one contains a factor $A(s) A(e) =0$. $\widetilde{W}_0$ vanishes by virtue of the second identity.

\section{Additional data for the one-dimensional Holstein polaron spectra\label{app:thermal_1d}}

Fig.~\ref{fig:thermal_1d_appendix} shows the spectral functions of the one-dimensional Holstein polaron discussed in Sec.~\ref{sec:temperature_dependence} over a wide temperature range, namely from $T=0.01$ to $T=10$. The model parameters are the same as in Fig.~\ref{fig:thermal_1d}. For the parameters studied here, the computational cost at high temperatures is comparable to that at low temperatures. Thus, our method offers broad temperature flexibility relative to DMRG and methods based on hierarchical equations of motion.

\bibliography{thermal_polaron}

\end{document}